# On the Role of Dispersion in One Model of Propagation of Elastic Excitations in Nerves

Alexander I.Kozlov [0000-0003-4650-9964]

Vitsebsk State Medical University, 27, Frunze str., Vitsebsk 210009, Belarus
e-mail: albapasserby@yahoo.com

**Abstract.** The nonlinear differential equation, which describes changes of membrane's density in the process of propagation of active potential along a nerve, had been studied. Depending on values of dispersion terms of that equation, forms and possible existence of exact periodic and solitary solutions were considered.

## 1. Introduction

The Hodgkin-Huxley ionic-channels model of propagation of excitations through biological membranes [1] seems to be the standard approach to description of motion of the active potential (AP) until now. However experimental results demonstrating change of mechanical properties of a membrane [2-3] and of (at least partial) reversibility of heat motion in the process of AP movement [4-6] are hardly to be understood in the framework of that model. Therefore different models of AP propagation as motion of nonlinear elastic disturbances along nervous tissues were proposed [7-10]. Detailed discussion of different approaches to the problem with voluminous lists of references could be found in works [11-13].

The present work is devoted to consideration of a so-called "soliton model" of AP propagation. This model has been suggested by Heimburg and Jackson [7] to include into consideration change of mass density as AP spreads along the surface of a membrane. (In fact the disturbances they considered do not exhibit all peculiar properties of solitons as it was also mentioned in paper [11], therefore we would call them "solitary displacements") Heimburg and Jackson proposed the next phenomenological nonlinear differential equation, which describes one-dimensional propagation of an elastic disturbance along a membrane [7]:

$$\frac{\partial^2 \Delta\rho}{\partial t^2} = \frac{\partial}{\partial x}\left\{\left[c_0^2 + p\cdot\Delta\rho + q\cdot(\Delta\rho)^2\right]\frac{\partial \Delta\rho}{\partial x}\right\} - h\frac{\partial^4 \Delta\rho}{\partial t^4} \tag{1}$$

where $\Delta\rho$ is the change of lateral mass density of a membrane in relation to the state of rest, $x$ and $t$ are space and time coordinates respectively, $c_0$ is the small-signal elastic velocity, $p$ and $q$ are coefficients of Taylor expansion of elastic compressibility (and thus of square of speed) on $\Delta\rho$. The last term in Eq.(1) containing a coefficient of proportionally $h$ is ad arbitrium introduced by authors to take into account dispersion of propagation velocity of elastic displacement, so to allow formation of solitary solutions to that equation.

Engelbrecht, Peets and Tamm [10] later amended Eq.(1) with addition of another one term containing the fourth-order mixed derivative on the right-hand side to take into account cylindrical profile of a nerve considered as a waveguide [10]. They also mentioned that insertion of such term allowed to avoid unbounded growth of propagation velocity at high frequencies [10]:

$$\frac{\partial^2 \Delta\rho}{\partial t^2} = \frac{\partial}{\partial x}\left\{\left[c_0^2 + p\cdot\Delta\rho + q\cdot(\Delta\rho)^2\right]\frac{\partial\Delta\rho}{\partial x}\right\} - h_1\frac{\partial^4\Delta\rho}{\partial x^4} + h_2\frac{\partial^4\Delta\rho}{\partial x^2\partial t^2} \tag{2}$$

Both coefficients $h_1$ and $h_2$ are believed to be positive, and the former one corresponds to the coefficient $h$ of Eq.(1).

Eqs. (1) and (2) were treated numerically [7, 8, 10] with resort to some analogy with the Boussinesq equation though there is a straightforward analytical method suitable for analysis of Eqs.(1)-(2). This method is used in the present communication in order to evaluate impact of dispersive coefficients $h_1$ and $h_2$ in Eq.(2) on its solutions.

The careful mathematical investigation of Eq.(1) (although its author mentioned the work of Engelbrecht *et al.*, but did not dwell on it) was also made in the paper [14] for different values of coefficients $p$, $q$, $h$ without paying attention to their measured values and physical existence and significance of results.

## 2. Solution

Using the standard group approach [15] it can be shown that Eqs. (1) and (2) have only trivial two-dimensional Lie algebra:

$$X_1 = \frac{\partial}{\partial t}, \qquad X_2 = \frac{\partial}{\partial x} \tag{3}$$

so that only the couple of incoming and outgoing disturbances are the possible invariants of those differential equations. Considering the incoming solution $z = x - vt$ like in the paper [7], one comes to the next fourth-order differential equation:

$$g\frac{\partial^4\Delta\rho}{\partial z^4} + \left(c_0^2 - v^2\right)\frac{\partial^2\Delta\rho}{\partial z^2} + p\frac{\partial}{\partial z}\left(\Delta\rho\cdot\frac{\partial\Delta\rho}{\partial z}\right) + q\frac{\partial}{\partial z}\left[(\Delta\rho)^2\cdot\frac{\partial\Delta\rho}{\partial z}\right] = 0 \tag{4}$$

where $g = v^2\cdot h_2 - h_1$.The latter ODE can be twice integrated successively giving:

$$g\frac{\partial^2\Delta\rho}{\partial z^2} + \left(c_0^2 - v^2\right)\Delta\rho + \frac{p}{2}(\Delta\rho)^2 + \frac{q}{3}(\Delta\rho)^3 + C_1 z + C_2 = 0 \tag{5}$$

where $C_1$ and $C_2$ are integration constants. The former constant $C_1$ is considered to be equal to zero to exclude solutions unbounded as $z$ tends to infinity. The independent as well as the dependent variables in Eq. (5) could be made dimensionless in the same way like that was proposed in the article [10]:

$$X = \frac{x}{l}, T = \frac{c_0 t}{l}, U = \frac{\Delta\rho}{\rho_0}, P = p\frac{\rho_0}{c_0^2}, Q = q\frac{\rho_0^2}{c_0^2}, G = \frac{g}{c_0^2 l^2}, V = \frac{v}{c_0}, Z = X - VT$$

where $l$ is a some characteristic length of the problem [10]. Then changing the dependent variable

$$U = \Phi - P/(2Q) \tag{6}$$

and assuming that

$$C_2 \equiv \left(1 - V^2 - \frac{P^2}{6Q}\right)\cdot\frac{P}{2Q} \tag{7}$$

one can obtain the Duffing equation in the next form:

$$\frac{d^2\Phi}{dZ^2} + \Phi\cdot\left(\frac{1-V^2}{G} - \frac{P^2}{4QG}\right) + \frac{Q}{3G}\cdot\Phi^3 = 0 \tag{8}$$

or

$$\frac{d^2\Phi}{dZ^2} + \alpha\cdot\Phi + \beta\cdot\Phi^3 = 0 \tag{9}$$

The next solutions are possible here depending on signs of coefficients $\alpha$ and $\beta$ (definition of those coefficients is evident from comparison of formulae (8) and (9)).

**Case I.** If $\alpha > 0$ and $\beta > 0$ then as it can be easily checked the next exact particular solution of Eq (9) being proportional to the Jacobi elliptic cosine function [16] is feasible:

$$\Phi(Z) = A\cdot \mathrm{cn}(u,k) \equiv A\cdot \mathrm{cn}(\lambda Z, k) \tag{10}$$

Here in (10) parameters $\lambda$ and $k$ depend on coefficients of Eq.(9) and on the amplitude $A$ of the solution, which in turn depends on initial conditions of the problem:

$$\lambda^2 = \alpha + \beta\cdot A^2 \quad ; \qquad k^2 = \frac{\beta\cdot A^2}{2\cdot(\alpha + \beta\cdot A^2)} \tag{11}$$

It is known that Jacobi elliptic functions of real variables have real values [16]. For both positive $\alpha$ and $\beta$ the values of $\lambda$ and $k$ are real and if $0 < k < 1$ then solutions (11)-(12) exhibit a periodic dependence $\Phi(Z)$ with the period equal to

$$T = 4K(k) \tag{12}$$

where $K(k)$ is the elliptic integral of the first kind [16].

**Case II.** On the other hand if $\alpha<0$ and $\beta>0$, but $\lambda^2$ remains positive and is very close to $\beta\times A^2/2$ so that $k\to 1$, then expression (10) would really tend to a solitary disturbance [16-17]:

$$\Phi(Z)\to A\cdot\mathrm{sech}(\lambda Z) \qquad (13)$$

which gives the next solution in initial dimensionless variables:

$$U(Z)=A\cdot\mathrm{sech}(\lambda Z)-\frac{P}{2Q} \qquad (14)$$

**Case III.** If $\alpha>0$ and $\beta<0$ then one can check that Jacobi elliptic sine [16] function satisfies Eq. (9):

$$\Phi(Z)=A\cdot\mathrm{sn}(u,k)\equiv A\cdot\mathrm{sn}(\lambda Z,k) \qquad (15)$$

where

$$\lambda^2=\alpha+\beta\cdot A^2 \quad ; \qquad k^2=-\frac{\beta\cdot A^2}{\alpha+\beta\cdot A^2} \qquad (16)$$

If $\alpha>-\beta A^2>0$, then real values of both $\lambda$ and $k$ are possible. Period of oscillation of solution (16) is also given by formula (12). In the limiting case $k\to 1$ the right-hand side of solution (15) leads to [16-17]:

$$\Phi(Z)\to A\cdot\tanh(\lambda Z) \qquad (17)$$

So the latter form.(17) of solution represents a weak-shock pulse.

## 3. Discussion

As it was stated experimentally [9] and used in previous calculations [7, 8, 10] for both considered kinds of biological membranes (unilamellar DPPC vesicles and bovine lung surfactants) measured numerical values of parameters in starting Eqs. (1), (2) could give negative $G$. The main reason of that is the negative sign of expression $h_2\cdot v^2-h_1$ in Eq.(2) or $h>0$ in Eq.(1) for both kinds of membranes whereas $q$ was measured to be positive. Therefore solutions (10)-(17) are invalid for initial equations because $\alpha<0$, $\beta<0$ and both parameters $\lambda$ and $k$ in elliptic functions (10) and (15) are imaginary in this case, so that questions about physical meaning of solutions arise.

It was formulated in one of early papers [8] devoted to the Heimburg-Jackson model (1) that "Although high frequency sound velocity measurements indicate that the dispersive coefficient, $h$, must be positive, neither the magnitude of $h$ nor the specific form of this term have been verified experimentally." Later Engelbrecht, Peets and Tamm also noted "the role and values of $h_1$ and $h_2$ need more explanation" [10].

## 4. Conclusion

Thus we should also conclude this communication with the statement that without better experimental investigation of values and signs of the dispersive terms in the Heimburg-Jackson model improved by Engelbrecht, Peets and Tamm (Eq.(2)) that model could hardly explain propagation of solitary elastic disturbances along nerve fibers. It seems to be that Eq.(2) with positive $h_2$ and $h_1 = 0$ is the simplest variant of the model allowing solitary as well as periodic solutions and non-contradicting the theory of wave propagation in elastic rods [18]:

$$\frac{\partial^2 \Delta\rho}{\partial t^2} = \frac{\partial}{\partial x}\left\{\left[c_0^2 + p \cdot \Delta\rho + q \cdot (\Delta\rho)^2\right]\frac{\partial \Delta\rho}{\partial x}\right\} + h_2 \frac{\partial^4 \Delta\rho}{\partial x^2 \partial t^2} \qquad (18)$$

## RERERENCES

1. A.L.Hodgkin, A.F.Huxley A quantitative description of membrane current and its application to conduction and excitation in nerve. *J. Physiol.* 117(4): 500-544,1952.
2. I.Tasaki, K.Iwasa. Rapid pressure changes and surface displacements in the squid giant axon associated with production of action potentials. *Jap. J. Physiol* 32(1): 69-81, 1982.
3. S.Terakawa. Potential-dependent variations of intracellular pressure in the intercellularly perfused squid giant axon. *J. Physiol.* 369(1), 229-248, 1986.
4. J.V.Howarth, R.D.Keynes, J.M.Ritchie. The origin of the initial heat associated with a single impulse in mammalian non-myelinated nerve fibres. *J. Physiol*: 194(3), 745-793, 1968.
5. I.Tasaki, P.M.Byrne. Heat production associated with a propagated impulse in bullfrog myelinated nerve fibers. *Jap. J. Physiol.*, 42(5): 805-813, 1992.
6. I.Tasaki, K.Kusano, P.M.Byrne. Rapid mechanical and thermal changes in the garfish olfactory nerve associated with a propagated impulse. *Biophys. J.*, 55(6):1033-1040, 1989.
7. T.Heimburg, A.D.Jackson. On soliton propagation in biomembranes and nerves. *Proc. Nat. Acad. Sci. USA*: 102(28), 9791-9795, 2005.
8. B.Lautrup, A.D.Jackson, T.Heimburg. The stability of solitons in biomembranes and nerves. ArXiv:physics/0510106v, 12 October 2005.
9. M.M.Rvachev. On axoplasmic pressure waves and their possible role in nerve impulses propagation. *Biophys. Rev. Lett.*: 5(2), 73-88 , 2010.
10. J.Engelbrecht, T.Peets, K.Tamm. Electromechanical coupling of waves in nerve fibres. *Biomech. Model. Mechanobiol.*: 17, 1771-1783, 2018.
11. H.Barz, A.Schreiber, U.Barz. Nerve impulse propagation: Mechanical wave model and HH model. *Med. Hypotheses*: 137(4) 109540, 2020.
12. C.Fillafer, A.Paeger, M.Schneider. The living state: How cellular excitability is controlled by the thermodynamic state of the membrane, *Progr. Biophys. Molec. Biol.*: 162(1), 57-68, 2021.

13. T.Heimburg. The mechanical properties of nerves, the size of the action potential, and consequences for the brain. *Chem. Phys. Lipids*: 267, 105461, 2025.
14. A.Elmandouh. Dynamical analysis of a soliton neuron model: bifurcations, quasi-periodic behaviour, chaotic patterns, and wave solutions. *Math*:, 13(12), 1912, 2025.
15. N.H.Ibragimov. *A Practical Course in Differential Equations and Mathematical Modelling*. ALGA Publications, Karlskrona, Sweden, 2008.
16. Yu.S.Sikorskii. *Introduction to the theory of elliptic functions with applications to mechanics*. Lenand Publisher, 2020, (in Russian).
17. A.P.Kuznetsov, S.P.Kuznetsov, N.M.Ryskin. *Nonlinear oscillations*. Lenand Publisher, 2020, (in Russian).
18. K.F.Graff. *Wave Motion in Elastic Solids*. Clarendon Press, 1975.